# StegTorrent: a Steganographic Method for the P2P File Sharing Service

Paweł Kopiczko, Wojciech Mazurczyk, Krzysztof Szczypiorski
Warsaw University of Technology, Institute of Telecommunications
Warsaw, Poland
e-mail: P.Kopiczko@stud.elka.pw.edu.pl, {wm, ksz}@tele.pw.edu.pl

*Abstract*—The paper proposes StegTorrent a new network steganographic method for the popular P2P file transfer service—BitTorrent. It is based on modifying the order of data packets in the peer-peer data exchange protocol. Unlike other existing steganographic methods that modify the packets' order it does not require any synchronization. Experimental results acquired from prototype implementation proved that it provides high steganographic bandwidth of up to 270 b/s while introducing little transmission distortion and providing difficult detectability.

## I. INTRODUCTION

BitTorrent [1], [2], a file-transfer system originally released in July 2001, is currently the most popular P2P (Peer-to-Peer) networking system worldwide [3] that allows users to share particular resources in the form of files. Studies show that the number of users exceeded 100 million in 2011, and that BitTorrent traffic accounts for about 94% of all P2P traffic, and is responsible for about 22% of North American fixed access daily traffic [4], which is continuously increasing each year. The success of BitTorrent primarily comes from two factors: its efficiency and openness. BitTorrent is significantly more efficient than classical client/server-based architectures. It allows peers sharing the same resource to form a P2P network, and then it focuses on fast and efficient replication to distribute the resource. It is also worth noting that because in BitTorrent a resource is divided into many fragments, a single peer is able to download many fragments simultaneously and it does not need the whole resource to share it. Additionally, BitTorrent software is free to download and many clients' versions are open source. This leads to the easy deployment of new applications and technologies, therefore stimulating further improvements.

Steganography encompasses all concealing techniques that embed a secret message into the carrier of this message in such a way that the carrier modification caused by the embedding of the steganogram must not be "noticeable" to anyone. It is important to emphasize that, for a third party observer who is not aware of the steganographic procedure, the exchange of steganograms remains hidden. Currently, one of the most dynamically evolving steganography subdisciplines is network steganography [10]. To perform hidden communication, it utilizes network protocols and/or relationships between them as the secret data carriers. Many methods were proposed and were aimed at specific network protocols from TCP/IP stack, for example, IP, TCP, UDP, and so on, or the whole service, for example, VoIP (Voice over IP) [11].

One of the most important aspects for every steganographic method is the choice of the proper carrier for the secret data. The most favorable carrier for secret messages must have two features:

- It should be popular, that is, usage of such a carrier should not be considered as an anomaly itself. The more such carriers are present and utilized in networks the better it is, because they mask the existence of hidden communication.
- Modification of the carrier related to the embedding of the steganogram should not be "visible" to the third party unaware of the steganographic procedure. Contrary to typical steganographic methods, which utilize digital media (pictures, audio and video files) as a cover for hidden data (steganogram), network steganography utilizes communication protocol control elements and their basic intrinsic functionality. As a result, such methods may be harder to detect and eliminate.

Thus, because of its popularity and traffic volume, BitTorrent's traffic is an ideal candidate for the hidden data carrier. In this paper, we present detailed analysis of potential opportunities for information hiding in BitTorrent.

Each network steganography method can be characterized by three features. First, steganographic bandwidth that describes how much secret data one is able to send using a particular method per time unit. Second, undetectability, defined as an inability to detect a steganogram inside certain carriers. The most popular way to detect a steganogram is to analyze the statistical properties of the captured data and compare them to the typical properties of that carrier. The last feature is the steganographic cost, which describes the degree of degradation of the carrier caused by the steganogram insertion procedure. The steganographic cost depends on the type of carrier, and, if it becomes excessive, it leads to easier detection of the steganographic method. For example, if the method uses voice packets as a hidden data carrier for steganographic purposes in IP telephony, then the steganographic cost is expressed in conversation degradation. If the hidden data carrier is a certain field of the protocol header, then the cost is expressed as a potential loss in that protocol functionality, and so forth.

For each network steganography method, a trade-off is always necessary between maximizing steganographic bandwidth and still remaining undetected (and retaining an acceptable level of the steganographic cost). A user can utilize a method naively and send as much secret data as possible, but it simultaneously raises the risk of disclosure. Therefore, he/she must purposely resign from some fraction of the steganographic bandwidth in order to be undetectable.

We wish to also emphasize that network steganography can be utilized by decent users to exchange covert data, for example, to circumvent censorship, to provide a communication channel between journalists and their information sources or by companies that are afraid of corporate espionage, but it can also be used by intruders to leak confidential data or to perform network attacks. This is the usual trade-off that requires consideration in a broader information hiding context, which is beyond the scope of this paper.

The rest of this paper is structured as follows. Section II introduces the basics of the BitTorrent P2P system. Section III presents the current state of research efforts on steganography in P2P networks and in BitTorrent in particular. Section IV describes the proposed steganographic method: StegTorrent, while in Section V the obtained experimental results are presented. Finally, Section VI concludes our work.

## II. BITTORRENT BASICS

BitTorrent is a P2P file sharing system that allows its users to distribute large amounts of data (especially large files) over IP networks. BitTorrent is distinguished from other similar file-transfer applications in that instead of downloading a resource (one or more files) from a single source (e.g. a central server), users download fragmented files from other users at the same time. As a result, the file-transfer time is considerably decreased because the group of users that share the same resource (or part of it) may consist of several to thousands of hosts. Such a group of users interested in the same resource (known as "peers") combine together with a central component (known as a "tracker") in BitTorrent. This combination of peers and trackers is called a "swarm". Trackers are responsible for controlling the resource transfer between the peers. Peers that hold onto a particular resource or part of a resource are required to share the resource and to perform the transfer.

We can distinguish two types of BitTorrent peers based on the stage at which they are involved in downloading or sharing a given resource. These types are:
- Seeds—peers that possess the complete resource and are only sharing it.
- Leechers—peers that do not possess the complete resource but they are interested in doing so. They also share the fragments they have already downloaded. When a leecher obtains all the remaining fragments of the resource it automatically becomes a seed.

In BitTorrent specification ([1], [2]), two main protocols are described that regulate data transfer: peer-tracker and peer-peer.

The connection between peer and tracker can be established with the use of an HTTP (Hypertext Transfer Protocol), or through UDP-based requests. Currently, the role of the tracker is diminished. The tracker is used mostly to initiate the connection with a swarm. After the connection is established, popular BitTorrent extensions like PEX (Peer Exchange) [2] or DHT (Distributed Hash Table) [4] are used. These extensions enable communication between peers without using a tracker as part of the swarm.

In a BitTorrent specification, *peer-peer data exchange* should be conducted using an application layer—a proprietary TCP-based protocol. It is a stateful protocol that is used to establish connections similar to the TCP handshake mechanism. However, it is also possible that instead of using TCP protocols in a transport layer, UDP-based μTP (μTorrent Transport Protocol) [15] should be used. This UDP-based protocol was introduced in 2009. It is not part of the original BitTorrent specification, but it was created by BitTorrent Inc., and as the results of a previous study indicated [14], it is currently the most popular choice.

The main aim of the μTP protocol is to efficiently manage usage of the available bandwidth during file transfers, while limiting the impact of file transfers on the on-going transmissions (especially non-BitTorrent related ones). The μTP protocol is capable of automatically reducing the rate at which BitTorrent packets are transmitted between peers in case there is interference with other applications running on the same host. This protocol uses a congestion control algorithm, which is a modified version of LEDBAT (Low Extra Delay Background Transport) [5], based on one-way delay measurements. The μTP protocol was implemented in the very popular BitTorrent client, μTorrent (http://www.utorrent.com) beginning with version 2.0, and it is currently used by default. This protocol is also used in other BitTorrent clients, such as in BitTorrent (http://www.bittorrent.com), Vuze (http://www.vuze.com), or Transmission (http://www.transmissionbt.com).

The μTP header format is illustrated in Fig. 1 and the field roles are explained below.

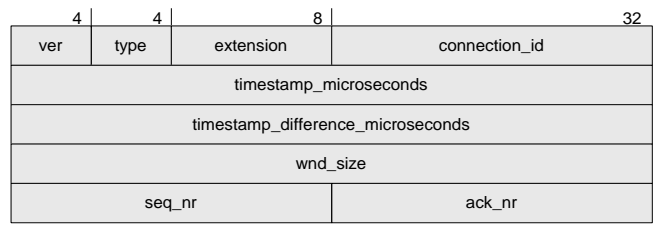

Figure 1. μTP header format.

The fields are utilized as follows [15]:
- *Ver*—version of the protocol. Currently version 1 is used.
- *Type*—packet type. Currently four types are defined: ST_DATA = 0—data packet, ST_FIN = 1— packet that ends the connection, ST_STATE = 2—acknowledgment packet (ack_nr field), without any other data, ST_RESET = 3—packet that forces the end of the connection, ST_SYN = 4—initializes the connection.
- *Extension*—type of the first extension on the extension list. Currently two types are specified: Selective acks and Extension. More about these extensions can be found in [15].

- *Connection_id*—random, unique number that is a connection identifier.
- *Timestamp_microseconds*—Number of microseconds from the last full second that passed from the time the last packet was sent from the local machine.
- *Timestamp_difference_microseconds*—the difference between the local time and the timestamp in the last received packet, at the time the last packet was received. This is the latest one-way delay measurement of the link from the remote peer to the local machine.
- *Wnd_size*—Advertised receive window (in bytes).
- *Seq_nr*—the sequence number of this packet.
- *Ack_nr*—the sequence number the sender of the packet last received from the remote peer.

From the abovementioned header fields, the *Timestamp_microseconds* would play an important role for the proposed steganographic method.

Typically, the BitTorrent client establishes many connections with other peers during the whole transmission process—typically about 70 [14]. BitTorrent connections are characterized by the high packet rate and a single resource is downloaded by a number of clients simultaneously (one-to-many transmission). This is also exemplarily illustrated in Figure 2 where it may be observed that the order in which packets to different clients are generated is highly variable and that the packets for different network localizations are heavily interleaved with each other.

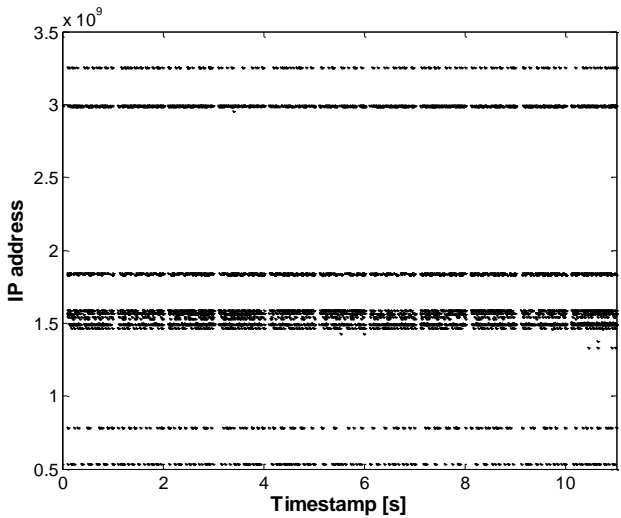

Figure 2. High variability in order in which packets to different clients are generated

That is why, due to its variability and high volume, the order in which the packets are sent or received by the client can be utilized as a carrier for hidden data exchange. This feature is used by StegTorrent to enable clandestine communication for the BitTorrent service.

### III. RELATED WORK

Steganography in P2P networks has been previously targeted by the research community. Information hiding techniques were utilized there to:
- Create a steganographic storage service—Mnemosyne [8]—that applies steganography from a local storage system [9] to a distributed, peer-to-peer system that is based on distributed hash tables,
- Build an adaptive covert communication system [12].

However, none of the abovementioned works proposed steganographic methods that utilize P2P-system-specific protocols—they reused well-known existing ones.

To the authors' best knowledge, only two papers discussed BitTorrent-specific steganographic methods. In 2008 Li et al. [6] proposed to utilize torrent meta files to hide information. Two approaches were specified there: one was based on changing the case of the letters in the URL addresses and the other reused some of the torrent file fields. Both these approaches are now obsolete since URLs are now being replaced by Magnet-URIs (http://magnet-uri.sourceforge.net).

Eidenbenz et al. [7] introduced an interesting concept for hidden communication in P2P networks and for BitTorrent in particular. The key idea is to allow conspiring peers to find one another without being revealed by means of the so called steganographic handshake and broadcast.

The authors propose enabling hidden communication to vary the fragment request sequence (request order channel). Few other methods are sketched there as alternatives for the above mentioned that is based on:
- Selection of a subset of fragments to be requested according to a shared secret—this is an enhancement of the method described above.
- Applying variations in the timing of the P2P protocol messages or the rate at which resource fragments are sent. For protocol messages it can be achieved, for example, by embedding secret data into the time between the reception of a fragment request and the corresponding transmission. For resource fragments (or rather packets), the rate at which they are sent is modified to enable hidden communication. For these methods, the authors conclude that they are feasible if the connection between the peers is stable and that, in real-life networks, usage of the correction codes is necessary. This is in fact a known problem for steganographic methods that utilize time relations [13] as well as requiring synchronization and achieving generally lower steganographic bandwidth than other methods.
- Setting the port number that contains hidden information—as the authors indicate themselves, this method is of limited use as a port number does not change during connection, thus only its initial value can be utilized. This results in very low steganographic bandwidth.

Unfortunately, it must also be emphasized that for the methods described above the authors did not provide any experimental results, thus they were not properly evaluated. However, they have proved feasible in the publicly available implementation BitThief (http://bitthief.ethz.ch).

In this paper, we propose a new steganographic method for BitTorrent—StegTorrent. The proposed method relies on the modification of the order of the data packets in the peer-peer data exchange protocol. The reordering of the packets is not new in the context of network steganography and it is a part of the broader group of methods that modify time relations between PDUs (Protocol Data Units) [10]. PDUs can be affected by modifying their inter-packet delay or introducing intentional losses or reordering. The first proposed utilization of the reordering for network protocol was by Kundur and Ahsan [18]. To be able to control packet sorting, they proposed using sequence number fields from the IPSec protocol headers: ESP (Encapsulating Security Payload) and AH (Authentication Header). Also Chakinala et al. [17] studied and formalized various models for hidden communication in ordered channels and presented simulation results for TCP traffic. However, all of these methods were lacking in undetectability or steganographic bandwidth. In real-life IP networks, the reordering in unicast transmissions (as the authors assume) is not witnessed often. Thus, when intentional reordering is exploited excessively it would be easy to discover, and when it is limited then the resulting steganographic bandwidth is low (typically a few bits per second or less). Moreover, such a solution typically requires a synchronization mechanism to correctly extract secret data. StegTorrent takes advantage of the facts that in BitTorrent there are usually one-to-many transmissions and that the µTP header provides a means for packet numbering and retrieval of their original sequence. This allows the provision of quite a high steganographic bandwidth (hundreds of bits per second) under the terms of undetectability with no requirements on synchronization.

## IV. STEGTORRENT DESCRIPTION

The clandestine communication scenario we consider for the proposed method is illustrated in Figure 3. We assume that both the secret data sending and receiving sides are in control of a certain number of BitTorrent clients and, as mentioned above, their IP addresses are known to each other. In Figure 3, for the sake of clarity, only single direction steganographic transmission is presented, but of course, end-to-end bidirectional communication is possible and the other direction is analogous. No knowledge of the network's topology is necessary. The hidden data sender uses the modified BitTorrent client—StegTorrent client—to share a resource that is downloaded by the second StegTorrent client that consists of a group of controlled BitTorrent clients.

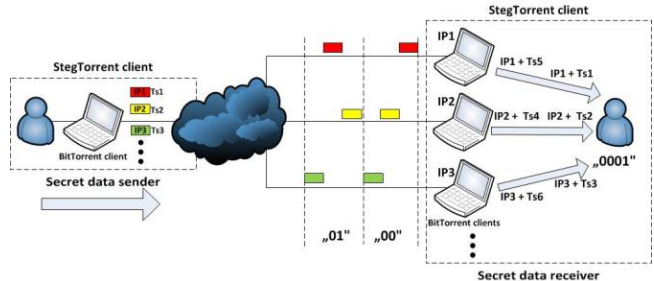

Figure 3. StegTorrent hidden data exchange scenario.

For the sake of the proposed method's description and analysis, we define the term *data package* as a set of IP addresses that is sent within the IP packets in a predetermined order and the term *data package size* as the total number of elements in this set. For example, let us assume that the data package size is 2. In this case, two packets with two different IP addresses (e.g. IP1 and IP2) are used to send bits of hidden data. In this simple scenario, if the order of the packets is modified for steganographic purposes, the BitTorrent client receives a packet that was sent from IP1 and then from IP2, then it will be interpreted as binary "0", in other case binary "1". We assume that the data package and its size are a shared secret between transmitting and receiving StegTorrent clients.

It must be noted that this method's performance depends on the size of the data package while the latter relies on the number of available receiving IP addresses (receiving BitTorrent clients under control).

### A. Sender side

As mentioned above, the secret data bits are encoded in the order in which the data packets are sent to the particular set of receiving clients. For example, if the secret data receiver controls two clients, A and B, then if the specific packet reaches A first and then B, then it would mean sending binary "0"; in the other case it will be interpreted as binary "1". To encode hidden data bits, Lehmer code is utilized [16] (the encoding is known a priori to the secret data sender and receiver). Then for $n$ different numbers of IP addresses we are able to encode $\lfloor \log_2(n!) \rfloor$ bits per data package. For example, for four packets with different IP addresses in a data package we are able to send 4 bits.

However, if the StegTorrent sender relies only on modifying the order of the data packets in which they were generated by the transmitter, then, due to poor network conditions, the order of the packet could be changed. This could potentially lead to problems with successful steganogram extraction. The solution is to incorporate into StegTorrent the intentional modification of the *timestamp_microseconds* field from the µTP protocol (see Section II). Values from these fields allow the unambiguous recognition of the order in which the packets were originally transmitted in the typical BitTorrent client.

By intentionally modulating these values while generating packets to enforce certain sequences, the sender can ensure that, even if the order of the transmitted packets is disrupted, the receiver would be able to correctly extract bits of the steganogram.

### B. Receiver side

The secret data receiver gathers all the information from the BitTorrent clients under the control of StegTorrent during the resource download session. Then it orders the packets' IP addresses based on their *timestamp_microseconds* values and begins the extraction of secret data bits.

To understand how the StegTorrent receiver works, the following example is presented below. Let us assume that the data package size is 5 and we denote packets sent to the five IP addresses: 1 when the packet is sent to IP1, 2 when the packet is sent to IP2 and so on. Let us further assume that by X we will be denoting packets sent to IP addresses outside the data package. If the exemplary packets' stream is:

1, 4, 4, X, X, 4, 4, 3, X, 3, 3, 5, 2, 2, 5, 5, 2, 2, 2, 4, 4, 4, X, 5, …

then, because the even number of packets from the same number are omitted and X are not taken into account, the receiver will extract the following secret bits colored with yellow:

**1**, 4, 4, X, X, 4, 4, **3**, X, 3, 3, **5**, 2, 2, 5, 5, 2, 2, **2**, 4, 4, **4**, X, | 5 …

The resulting data package is {1, 3, 5, 2, 4}. Then this package is mapped to certain predefined secret data bit sequences. The last packet from IP5 (marked with grey) is the beginning of the new data package since the size of the data package is 5.

If some packets are lost then the secret data receiver will wait for retransmission which is ensured by the μTP protocol. The retransmitted packet will have the same *timestamp_microseconds* value as the original one. After the packet is retransmitted and the received data is complete, the secret data receiver is ready to extract the secret data.

## V. EXPERIMENTAL RESULTS

To experimentally evaluate the proposed steganographic method, the StegTorrent prototype implementation was developed and the test-bed was setup (Fig. 4). The presented scenario is a simpler version of that presented in Figure 3; however, the main principles and functioning of the proposed method remain the same. In this test-bed, twenty sessions of resource downloading were performed and the average results are presented. For each session 500 000 μTP packets and the corresponding logs were captured and subjected to analysis.

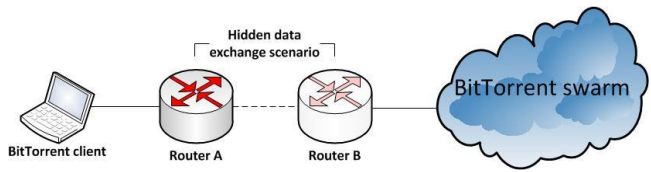

Figure 4. StegTorrent hidden data exchange scenario

First, we analyzed the number of packets that were sent from different IP addresses during each session. The information obtained can then be utilized to pinpoint the best number of IP addresses that can be used to create a package and to achieve the highest steganographic bandwidth under the terms of the method's undetectability.

The obtained results are presented in Figure 5. Only the ten most active IP addresses are presented separately and the rest are aggregated. Each session is described as *xTy* where x denotes the number of downloaded resources and y is the number of the session. As expected, the results vary greatly, which is beneficial from the undetectability point of view. That is why, in Figure 6, we also present the most different cases as well as the average.

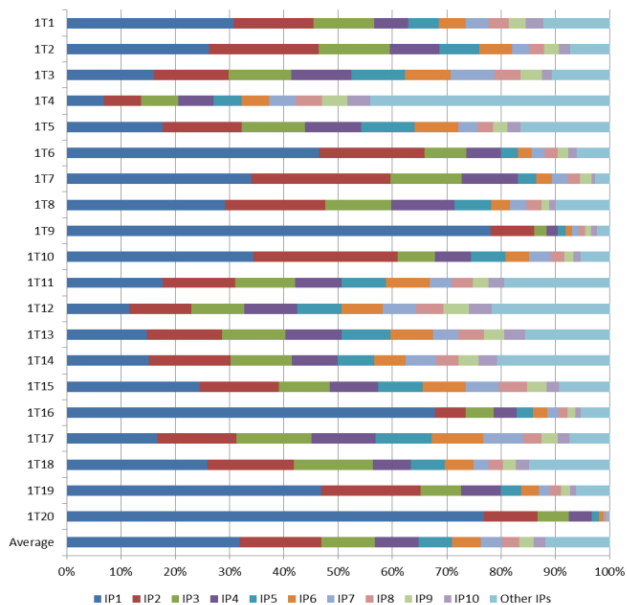

Figure 5. Distribution of number of packets sent from different IP addresses during single resource downloading session (in %); xTy – x denotes number of downloaded resources and y is the number of the session.

It turned out that in every session there was at least one IP address that was dominant and responsible for sending, in most cases, more than 25% of the total number of packets (the result falls in the 16–77% range). It is also worth noting that, in most cases, more than 75% of all packets are sent by six different IP addresses and that this is why this number was selected as the maximum size of the analyzed package in the following measurements.

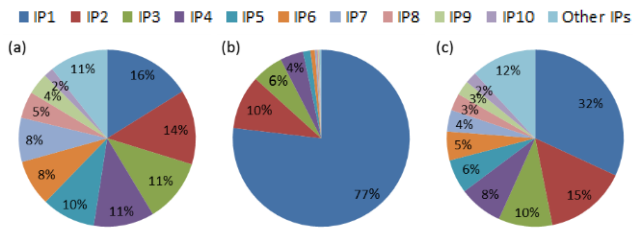

Figure 6. Examples of distribution of number of packets sent from different IP addresses during a single resource downloading session (in %) – (a) balanced (1T3), (b) single IP address dominant (1T20), (c) average.

Next, we analyzed the achieved StegTorrent steganographic bandwidth as well as the *utilization of packets* in the created covert channel that is defined as the ratio between the number of packets used for the creation of data packages (thus sending secret data) and number of all packets received during the session. The measurements were conducted for three cases, where:
- All of the IP addresses are utilized for data package creation (Case A),
- The number of IP addresses equals the size of the package (Case B),
- The number of IP addresses used for data package creation is 6 (Case C).

Case A is the least practical due to the requirement to control the endpoints with all IP addresses that took part in resource downloading—Cases B and C are definitely more realistic. Moreover, for B and C only, the most active IP addresses were utilized, that is those that sent the largest number of packets during the downloading session. In each case, the five different sizes of data packages were analyzed from 2 to 6. The following figures and tables present the average results.

Table I presents the obtained experimental results for single resource downloading sessions for cases A–C. Obviously, the best results were achieved for Case A where all IP addresses were used for package creation. However, as mentioned before, in practice, this is the most difficult scenario to accomplish. Steganographic bandwidth for this case is in the range 430–660 b/s and it increases with the increase in package size. On the other hand, packet utilization with small package sizes is more than 50% while, with the package size increase, it falls to about 30%.

TABLE I. EXPERIMENTAL RESULTS FOR SINGLE RESOURCE DOWNLOADING SCENARIO

| Case # | Data package size | Steganographic bandwidth [b/s] | Standard deviation [b/s] | Packets utililzation [%] |
|---|---|---|---|---|
| A | 2 | 438.41 | 106.43 | 65.05 |
| A | 3 | 467.28 | 120.51 | 51.69 |
| A | 4 | 590.49 | 162.88 | 43.36 |
| A | 5 | 607.73 | 176.89 | 37.07 |
| A | 6 | 656.01 | 200.88 | 31.93 |
| B | 2 | 82.82 | 54.87 | 11.50 |
| B | 3 | 66.29 | 38.33 | 6.95 |
| B | 4 | 67.99 | 52.16 | 4.24 |
| B | 5 | 63.99 | 46.10 | 2.99 |
| B | 6 | 74.30 | 51.25 | 2.51 |
| C | 2 | 273.00 | 93.27 | 40.49 |
| C | 3 | 239.02 | 91.96 | 26.17 |
| C | 4 | 227.95 | 106.01 | 16.46 |
| C | 5 | 138.59 | 96.52 | 8.10 |
| C | 6 | 74.30 | 51.25 | 2.51 |

The experimental results obtained for Case C were significantly better than for Case B for small package sizes. It is worth noting that the steganographic bandwidth for Case B and a package size of 6 is more than three times lower than for Case C and package size 2, while the latter is easier to implement. This is caused by the usually not well balanced distribution of packets from different IP addresses (Fig. 6, c). That is why, when the larger data package size is utilized, a considerable number of packets from IP addresses that are chosen for steganographic purposes are not carrying any secret data (they are not included in packages). This in turn results in very low packet utilization for larger data package sizes (<10% for sizes equal to 5 or 6).

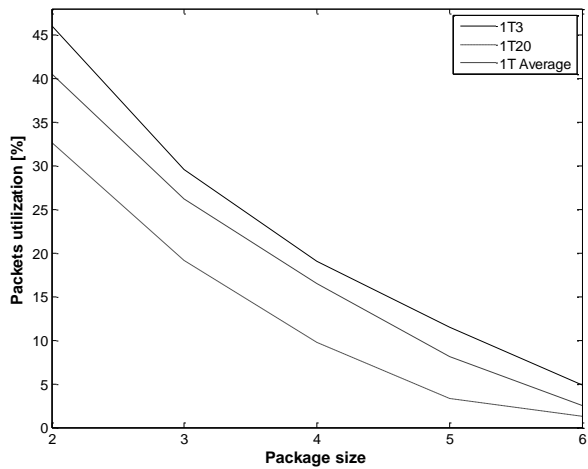

Figure 7. Packet utilization for different data packages sizes (case C).

Finally, Figure 7 illustrates packet utilization for different data packages sizes for Case C and sessions 1T3, 1T20 and the average (from Fig. 6). It proves that packet utilization in StegTorrent is related to the distribution of the number of packets sent from different IP addresses, but not as much as could be expected. In other words even if there is a dominant IP address (like in Fig. 6, b) packet utilization does not decrease considerably when compared to the average results (about 7%) and the difference is smaller for larger data package sizes (2–5%). Also, as mentioned before, packet utilization decreases with the increase in data package size.

The StegTorrent steganographic cost is negligible, since it only introduces a small delay in resource downloading by re-sorting the packets in the predefined order. This is completely transparent to the BitTorrent client.

## VI. CONCLUSIONS

This paper presents a new BitTorrent-specific method called StegTorrent that relies on the modification of the order of data packets in the peer-peer data exchange protocol. StegTorrent takes advantage of the facts that in BitTorrent there are usually many-to-one transmissions and that the μTP header provides a means of packet numbering and retrieval of their original sequence. This allows the provision of quite a high steganographic bandwidth, of about 270 b/s, for the most realistic scenario under the terms of undetectability and with no requirement for synchronization.

ACKNOWLEDGMENT

This research was partially supported by the Polish Ministry of Science and Higher Education and the Polish National Science Center under grants: 0349/IP2/2011/71 and 2011/01/D/ST7/05054.